\documentclass[journal]{IEEEtran}
\usepackage{cleveref}
\usepackage[usenames,dvipsnames,svgnames,table]{xcolor}

\usepackage{cite}
\ifCLASSINFOpdf
\else
\fi
\usepackage{amsmath}
\usepackage{amssymb}
\usepackage{mathtools}
\usepackage{bbm}
\usepackage{bm}

\usepackage[linesnumbered,nosemicolon]{algorithm2e}

\usepackage{stfloats}

\newcommand{\prob}[1]{\text{p}({#1})}
\newcommand{\probhat}[1]{\widehat{\text{p}}\left({#1}\right)}
\newcommand{\prop}[2]{\text{q}\left({#1|#2}\right)}
\newcommand{\proposalf}{\text{q}}

\newcommand{\Np}{\text{N}_\text{p}}

\newcommand{\transpose}{^\text{T}}

\newcommand{\state}{\mathbf{s}}
\newcommand{\obs}{\mathbf{z}}
\newcommand{\nmcruns}{\text{N}_\text{MC}}

\newcommand{\ContinuousUniform}[2]{\text{Unif}\left({#1,#2}\right)}

\newcommand{\tk}{\text{k}}

\newcommand{\CurrentState}{\state_{\tk}^{m}}
\newcommand{\ProposedState}{\state_{\tk}^{*}}
\newcommand{\NewState}{\state_\tk^{m+1}}

\newcommand{\AcceptanceRatio}{\mathcal{A}\left(\CurrentState,\ProposedState\right)}

\newcommand{\obss}{\text{z}}
\newcommand{\Nbi}{\text{N}_\text{bi}} % number of burn-in iterations
\renewcommand{\CurrentState}{\state_{\tk}^{i}}
\renewcommand{\NewState}{\state_{\tk}^{i+1}}
\newcommand{\AcceptanceRatioDefLkl}{\AcceptanceRatio = \min\left\{1, \frac{\prob{\obs_{\tk}|\state_{\tk}^{*}}}{\prob{\obs_{\tk}|\state_{\tk}^{i}}}\right\}}
\newcommand{\AcceptanceRatioRef}{\mathcal{A}\left(\NewState,\ProposedState\right)}

\newcommand{\AcceptanceRatioDefRL}{\AcceptanceRatioRef = \min\left\{1, \frac{ \prob{\obs_{\tk}|\state_{\tk}^{*}} \prob{\state_{\tk}^{*}|\state_{\tk-1}^{i+1}} }{ \prob{\obs_{\tk}|\state_{\tk}^{i+1}} \prob{\state_{\tk}^{i+1}|\state_{\tk-1}^{i+1}} } \frac{ \text{q}_\text{RL}\left({\state_{\tk}^{i+1}|\state_{\tk}^{*}, \state_{\tk-1}^{i+1}, \obs_{\tk}}\right) }{ \text{q}_\text{RL}\left({\state_{\tk}^{*}|\state_{\tk}^{i+1}, \state_{\tk-1}^{i+1}, \obs_{\tk}}\right) } \right\}}

\newcommand{\xvec}{\mathbf{x}}
\newcommand{\alphavec}{\bm{\alpha}}
\newcommand{\muvec}{\bm{\mu}}

\newcommand{\mvnpdff}{\phi}
\newcommand{\mvnpdf}[1]{\mvnpdff\left(#1\right)}

\newcommand{\Fisherf}{\text{F}}
\newcommand{\Fisher}[1]{\Fisherf\left(#1\right)}

\newcommand{\tA}{\text{A}}

\newcommand{\tJ}{\text{J}}

\renewcommand{\th}{\text{h}}

\begin{document}
\title{Riemann-Langevin Particle Filtering\\ in Track-Before-Detect}

%\author{Michael~Shell,~\IEEEmembership{Member,~IEEE,}
%        John~Doe,~\IEEEmembership{Fellow,~OSA,}
%        and~Jane~Doe,~\IEEEmembership{Life~Fellow,~IEEE}% <-this % stops a space
\author{Fernando J. Iglesias-Garcia,
        Pranab K. Mandal,
        M\'{e}lanie Bocquel,
        Antonio G. Marques% <-this % stops a space
\thanks{Work in this paper was partially supported by the Spanish MINECO grants No TEC2013-
	41604-R and TEC2016-75361-R, and the TRAX project from the Seventh Framework Programme
	under grant agreement N\textordmasculine $607400$.}
\thanks{F. Iglesias-Garcia and A. G. Marques are with the Department
of Signal Theory and Communications, King Juan Carlos University.}% <-this % stops a space
\thanks{M. Bocquel and P. K. Mandal are with the Faculty of Electrical Engineering, Mathematics, and Computer Science, University of Twente.}% <-this % stops a space
\thanks{Manuscript received Month Day, 2017; revised Month Day, 2017.}}


\markboth{IEEE SIGNAL PROCESSING LETTERS,~Vol.~X, No.~X, Month~2017}%
{IEEE SIGNAL PROCESSING LETTERS,~Vol.~X, No.~X, Month~2017}



\maketitle

\begin{abstract}
Track-before-detect (TBD) is a powerful approach that consists in providing the tracker with sensor measurements directly without pre-detection. Due to the measurement model non-linearities, online state estimation in TBD is most commonly solved via particle filtering. Existing particle filters for TBD do not incorporate measurement information in their proposal distribution. The Langevin Monte Carlo (LMC) is a sampling method whose proposal is able to exploit all available knowledge of the posterior (that is, both prior and measurement information). This letter synthesizes recent advances in LMC-based filtering to describe the Riemann-Langevin particle filter and introduces its novel application to TBD. The benefits of our approach are illustrated in a challenging low-noise scenario.
\end{abstract}


% \fi
\IEEEpeerreviewmaketitle

\section{Introduction}\label{sec:intro} 

\IEEEPARstart{S}{equential state} estimation in nonlinear dynamical systems is a challenging problem. Following a Bayesian approach, a closed-form expression of the conditional probability of the state (posterior) is only attainable for a restricted class of models. Therefore, methods based on numerical approximations are oftentimes employed. Among these approximations, Monte Carlo (MC) methods \cite{gordon1993novel,doucet2001sequential,arulampalam2002tutorial} are popular due to their flexibility and provable convergence guarantees. %provided that enough computational time and power are available.
Particle filters (PFs) based on importance sampling (IS) are a straightforward implementation of MC methods in dynamical systems. However, their practical application is quickly challenged as the dimension of the state space increases. Moreover, IS-based PF usually require resampling in order to avoid sample degeneracy, restraining their parallel implementation. Due to these limitations, the integration of Markov chain Monte Carlo (MCMC) in PF has emerged as competing alternative \cite{khan2005mcmc,smith2008tracking,pang2011detection,bocquel2013multitarget}.

This paper investigates the application of a certain type of MCMC-based PF to the problem of track-before-detect (TBD) \cite{salmond2001particle,boers2004multitarget,rutten2005recursive}. TBD consists in removing the detection stage typically found before the tracking module in surveillance and tracking systems. Consequently, the ``unthresholded'' image measurements are fed directly into the tracker. Due to the non-linearities in TBD measurement models, PFs are commonly employed. Nevertheless, the proposal distributions in PFs (based on either MCMC or IS) developed for TBD have solely exploited the prior. This is in general inefficient, but especially when most of the posterior information is contained in the measurement. In this context, the Langevin Monte Carlo (LMC), which is a MCMC whose proposal exploits measurement knowledge through the Langevin equation \cite{duane1987}, emerges as an efficient alternative that has been recently applied in the context of classical tracking \cite{iglesias2015langevin}.

The main goal of this letter is the application of LMC-based PFs to TBD. Together with the particularization to the TBD setup, we also present several modifications to enhance the original LMC filtering in \cite{iglesias2015langevin}. One of them is related to the fact of LMC being parameterized by a step size that must be carefully adjusted for every time step. Ideally, an LMC-based filter should systematically \emph{adapt} its step size. Here, this problem is avoided employing the Riemannian MCMC \cite{girolami2011riemann}. Furthermore, the filter in \cite{iglesias2015langevin} approximates the gradient of the posterior and uses an MCMC-based PF methodology that introduces a bias \cite{iglesias2015acceptance}. In this letter these limitations are addressed using the \textit{sequential} MCMC approach \cite{pang2011detection,septier2016langevin}.% we keep the standard acceptance probability expression and still retain the computational cost low following the sequential MCMC framework.

The letter is organized as follows. \Cref{sec:smcmc} discusses fundamentals of MCMC-based PFs and sequential MCMC, with a special focus on the acceptance probability. The Langevin proposal and its extension based on differential geometry are described in \Cref{sec:langevin_proposal}. The Fisher information matrix \cite{amari1985differential}, plays a key role on this extension. Hence, \Cref{sec:tracklkl_fishermet} derives the Fisher information matrix for non-linear Gaussian models and \Cref{sec:tbd} particularizes it for TBD. \Cref{sec:rlmcf} details the Riemann-Langevin MC filter. The performance of the filter is analyzed in a original low-noise TBD application in \Cref{sec:experiment}. The conclusions in \Cref{sec:conclusion} close the letter.

\section{Preliminaries: Sequential MCMC}
\label{sec:smcmc}

This section introduces notations to be used in the letter and illustrates how sequential MCMC renders an efficient MCMC-based PF. We first introduce the general expression for the probability of accepting a sample, and then show how this expression simplifies for sequential MCMC. These methods were introduced in a tracking application \cite{pang2011detection}, with the goal of enabling efficient computation of the acceptance probability when performing block sampling (e.g. sampling one object at-a-time). The gains in sequential MCMC come from sampling the \emph{joint} posterior at two consecutive time steps. %Sampling jointly at two consecutive time steps (denoted by $\tk$ and $\tk-1$) is efficient because it makes the acceptance probability not dependent of the prediction.

%As discussed earlier in \Cref{sec:langevin_proposal}, by using the proposal in \Cref{eq:proposal_rl} it becomes possible to sample from high-dimensional probability distributions of the state without the need of resorting to a block sampling strategy. In this section we show that the sequential MCMC framework should still be used with the proposal in \Cref{eq:proposal_rl} even if block sampling is not necessary.

To be specific, let the state and measurement vectors at time step $\tk$ be denoted by $\state_{\tk}$ and $\obs_{\tk}$, respectively, and let $\obs_{1:\tk}$ denote the sequence of measurements $\obs_{1},\dots,\obs_{\tk}$.  Furthermore, the Bayesian posterior is $\prob{\state_\tk|\obs_{1:\tk}}$, the prediction $\prob{\state_{\tk}|\obs_{1:\tk-1}}$, and the measurement likelihood $\prob{\obs_{\tk}|\state_{\tk}}$. Suppose that an MCMC algorithm is used to sample from $\prob{\state_\tk|\obs_{1:\tk}}$. Let the current sample of the Markov chain be denoted by $\state_\tk^i$ and the proposed sample by $\state_\tk^*$, drawn from the proposal distribution $\proposalf$. A key component of a MCMC is to decide if $\state_\tk^*$ is accepted or rejected. The general expression for the acceptance probability in the Metropolis-Hastings algorithm involves the ratios of the posteriors and the proposals and is given by
\begin{equation}
\min\left\{1,\frac{\prob{\obs_\tk|\state_\tk^*} \prob{\state_\tk^*|\obs_{1:\tk-1}}}{\prob{\obs_\tk|\state_\tk^i} \prob{\state_\tk^i|\obs_{1:\tk-1}}} \frac{\prop{\state_\tk^i}{\state_\tk^*}}{\prop{\state_\tk^*}{\state_\tk^i}}\right\}.
\label{eq:accprob}
\end{equation}
%
%In the expression above the posterior has been expanded into the product given by Baye's theorem: $\prob{\state_\tk|\obs_{1:\tk}} \propto \prob{\obs_\tk|\state_\tk} \ \prob{\state_\tk|\obs_{1:\tk-1}}$. The normaliser $\prob{\obs_\tk|\obs_{1:\tk-1}}$ is simplified as it appears in the numerator and denominator stemming from $\prob{\state_\tk^*|\obs_{1:\tk}}$ and $\prob{\state_\tk^i|\obs_{1:\tk}}$, respectively.
%
Clearly, unless the proposal is simply equal to the prediction, computing \eqref{eq:accprob} requires evaluating the density $\prob{\state_\tk|\obs_{1:\tk-1}}$.%Note that we have dropped the conditional in the proposal since for this choice of proposal there is no dependence on the current state of the Markov chain.% Even though making the acceptance probability simple, this choice of proposal usually results in an inefficient MCMC algorithm \cite{septier2016langevin}. % \textcolor{red}{TODO add reference to Bickel if their analysis also uses prediction as proposal}, especially in high-dimensional problems such as multiple object tracking.

The sequential MCMC brings flexibility in choosing the proposal while maintaining efficient computation of the acceptance probability. The key is to consider jointly the state at $\tk$ and $\tk-1$. Within each time step of the PF, an MCMC samples from the joint density $\prob{\state_\tk,\state_{\tk-1}|\obs_{1:\tk}}$. Due to the factorization of the joint density,
\begin{equation}
\prob{\state_\tk,\state_{\tk-1}|\obs_{1:\tk}} \propto \prob{\obs_\tk|\state_\tk} \prob{\state_\tk|\state_{\tk-1}} \prob{\state_{\tk-1}|\obs_{1:\tk-1}},
\label{eq:joint_expansion}
\end{equation}
a sample from a proposal $\proposalf$ on $\state_\tk$ is accepted with probability
\begin{equation}
\min\left\{1,\frac{\prob{\obs_\tk|\state_\tk^*} \prob{\state_\tk^*|\state_{\tk-1}^i}}{\prob{\obs_\tk|\state_\tk^i} \prob{\state_\tk^i|\state_{\tk-1}^i}} \frac{\prop{\state_\tk^i}{\state_\tk^*,\state_{\tk-1}^i}}{\prop{\state_\tk^*}{\state_\tk^i,\state_{\tk-1}^i}}\right\}.
\label{eq:accprob_smcmc}
\end{equation}
Note that the factors corresponding to $\prob{\state_{\tk-1}|\obs_{1:\tk-1}}$ in \eqref{eq:joint_expansion} simplify. Compared to \eqref{eq:accprob}, the sequential MCMC's acceptance probability \eqref{eq:accprob_smcmc} depends on the evaluation of the transition model $\prob{\state_\tk|\state_{\tk-1}}$ in lieu of the prediction $\prob{\state_{\tk}|\obs_{1:\tk-1}}$.

%Similarly, if the proposal is not the prediction and it samples all the variables in $\state_\tk$ the acceptance probability in \eqref{eq:accprob} requires evaluations of the prediction (\eqref{sec:langevin_proposal} describes an example of this type of proposal). Hence, the sequential MCMC is also beneficial even if a block sampling strategy is not employed.

\newcommand{\textcite}{\cite}
\newcommand{\parencite}{\cite}
\section{Riemann-Langevin proposal in Track-Before-Detect}
\label{sec:langevin_proposal}

This section introduces the application of the LMC filter to tracking applications and presents the enhancements described in \Cref{sec:intro}. But first, let us review the generalization of the Langevin proposal inspired by differential geometry (namely, the Riemann-Langevin proposal) \cite{girolami2011riemann}.

We start with some notational conventions: $(\state_{\tk},\state_{\tk-1})$ is the current sample or state of the Markov chain, $\mathcal{N}\left(\xvec; \muvec, \Sigma\right)$ is a Gaussian density with mean $\muvec$ and covariance $\Sigma$, $\nabla_{\xvec} \ \text{f}(\xvec)$ is the gradient of $\text{f}$ w.r.t. $\xvec$, and $\Fisherf$ is a metric tensor representing the Fisher information matrix. With these conventions, a proposed sample $\state_{\tk}'$ is obtained from the Riemann-Langevin proposal as given in \eqref{eq:proposal_rl} at the bottom of the page.

\begin{figure*}[b]
\begin{equation}
\text{q}_{\text{RL}}\left({\state_{\tk}'|\state_{\tk},\state_{\tk-1},\obs_{\tk}}\right) = \mathcal{N}\left(\state_{\tk}'; \state_{\tk} + \frac{\epsilon^2}{2} \Fisherf^{-1}\left(\state_{\tk},\state_{\tk-1},\obs_{\tk}\right) \nabla_{\state_{\tk}} \log \left(\prob{\obs_{\tk}|\state_{\tk}} \ \prob{\state_{\tk}|\state_{\tk-1}} \right), \epsilon^2 \Fisherf^{-1}\left(\state_{\tk},\state_{\tk-1},\obs_{\tk}\right)\right)
\label{eq:proposal_rl}
\end{equation}
\end{figure*}

This proposal leverages the property that probability distributions lie in Riemann manifolds and stems from extending the Langevin equation with the metric tensor \textcite{girolami2011riemann}. A sensible choice of the metric tensor containing curvature information about the posterior is the Fisher information matrix \parencite{amari1985differential}. In sequential MCMC it is defined as \cite{septier2016langevin}
\begin{equation}
\Fisher{\state_{\tk},\state_{\tk-1},\obs_{\tk}} \!=\! -\mathbb{E}_{\obs_{\tk}|\state_{\tk}} \!\left[\Delta_{\state_{\tk}}^{\state_{\tk}} \log \left(\prob{\obs_{\tk}|\state_{\tk}}  \prob{\state_{\tk}|\state_{\tk-1}}\right)\right],\!
\label{eq:tensor_metric}
\end{equation}
where $\mathbb{E}_{\obs_{\tk}|\state_{\tk}}$ denotes the expected value w.r.t. $\prob{\obs_{\tk}|\state_{\tk}}$ and $\Delta_{\xvec}^{\xvec} \ \text{f}(\xvec)$ the Hessian of $\text{f}$ w.r.t. $\xvec$.% The matrix $\Fisherf$ is positive-definiteness providing that $\prob{\state_{\tk}|\state_{\tk-1}}$ is log-concave.

%The proposal in \eqref{eq:proposal_rl} takes into account different scalings of the variables and correlations via $\tG$. Consequently, it becomes possible to perform high-dimensional filtering without resorting to block or Metropolis-within-Gibbs sampling. This is particularly useful for cases where the variables of the state space are highly correlated.

\subsection{Information matrices in tracking applications}
\label{sec:tracklkl_fishermet}

Sampling from \eqref{eq:proposal_rl} requires the information matrix of the product $\prob{\obs_{\tk}|\state_{\tk}} \prob{\state_{\tk}|\state_{\tk-1}}$. If the models are Gaussian and linear, the information matrix is simply the covariance matrix. However, important tracking applications such as TBD have non-linear $\prob{\obs_{\tk}|\state_{\tk}}$. Therefore, this section reviews the information matrix of a non-linear multi-variate Gaussian. The next section will build upon this result and provide the information matrix for TBD.

Consider a Gaussian density function $\mvnpdff$ with covariance $\Sigma$ and mean $\muvec$. Suppose that $\muvec$ is a function of a vector $\alphavec$ and that $\Sigma$ is independent of $\alphavec$. Then, the gradient and information matrix w.r.t $\alphavec$ are \cite{kay1993fundamentals}:
\begin{align}
\nabla_{\alphavec} \log \mvnpdf{\xvec} &= \left(\frac{\partial \muvec}{\partial \alphavec}\right)\transpose \Sigma^{-1} \left(\xvec-\muvec\right) \label{eq:mvngradient} \\
\Fisher{\alphavec} &= \left(\frac{\partial \muvec}{\partial \alphavec}\right)\transpose \Sigma^{-1} \left(\frac{\partial \muvec}{\partial \alphavec}\right) \label{eq:mvnfisher}.
\end{align}

\subsection{Track-before-detect}
\label{sec:tbd}
In this section we provide the gradient and information matrix of the TBD measurement model in \cite{boers2004multitarget}. 
%To this end, the section starts reviewing the TBD model.
\subsubsection{Measurement model}
\noindent
Consider an imaging sensor (e.g. camera, radar) that measures the vector $\obs_{\tk}$ composed of the scalars $\obss_{\tk}^{(j)}$ in $\tJ$ pixels or cells:
\begin{equation}
\obs_{\tk} = [\obss_{\tk}^{(1)} \ \cdots \ \obss_{\tk}^{(\tJ)}]\transpose. 
\end{equation}

\begin{figure}
\centering
\includegraphics[width=\columnwidth]{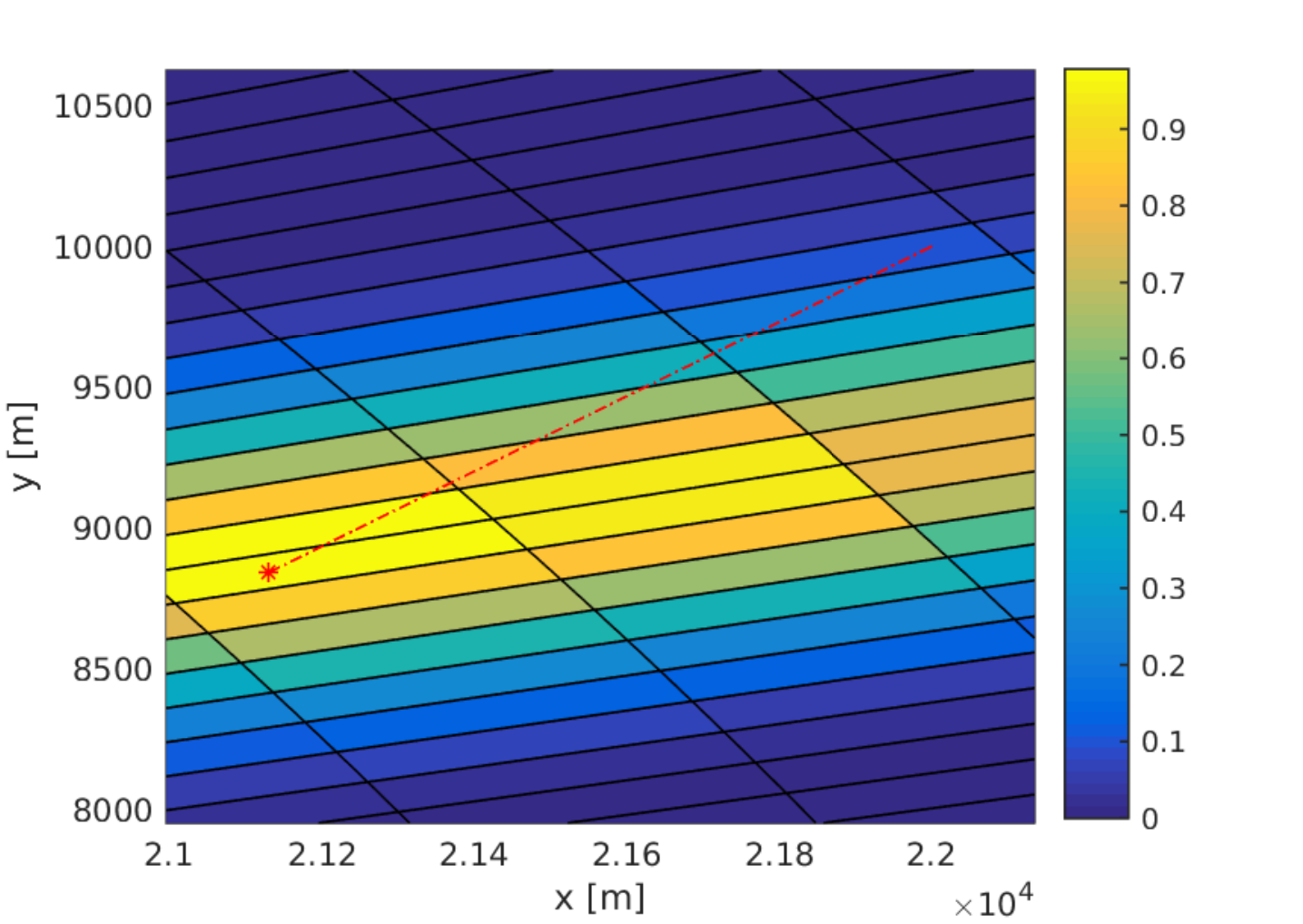}
\caption{Range-bearing imaging sensor grid, measurement, and object trajectory (dotted line) illustrating the TBD model. The solid lines depict the boundaries of the cells. The colors of the cells denote the measurements \eqref{eq:meas_model}.}
\label{fig:sensor_measurement_trajectory}
\end{figure}

Suppose that the strength of the measured signal under the influence of an object is constant and denoted by $\tA$. In addition, let the measurement noise $w_{\tk}^{(j)}$ be zero-mean Gaussian with variance $\sigma_w^2$, so that the signal-to-noise ratio (SNR) is $20 \log_{10} \left( \tA/\sigma_w\right)$. Then, the measurement in a single cell is
\begin{equation}
% WITH CUT-OFF PSF.
%\obss^{(j)} = \widehat{\obss}^{(j)} + w^{(j)} = \mathbbm{1}_{\tC(\state)}(j) \ \tA \ \th^{(j)}(\state) + w^{(j)},
% WITHOUT CUT-OFF PSF.
\obss_{\tk}^{(j)} = \widehat{\obss}_{\tk}^{(j)} + w_{\tk}^{(j)} = \tA \ \th^{(j)}(\state_\tk) + w_{\tk}^{(j)},
\label{eq:meas_model}
\end{equation}
where $\text{h}^{(j)}$ is the \textit{point spread function} characteristic of the sensor. While our results hold for any point spread function, a popular choice (also used for the experiments in \Cref{sec:experiment}) is \cite{boers2004multitarget}
\begin{equation}
\text{h}^{(j)}(\state_\tk) = \exp \left(-\frac{(r_j-r(\state_\tk))^2}{2\text{R}}-\frac{(b_j-b(\state_\tk))^2}{2\text{B}} \right),
\end{equation}
where $r_j$ and $b_j$ denote, respectively, the range and bearing cell centroids; $r(\state_\tk)$ and $b(\state_\tk)$ the position in polar coordinates; and \text{R} and \text{B} are sensing parameters that depend on the resolution and the boundaries of the surveillance area \cite{boers2004multitarget}. Other measurement dimensions (e.g. Doppler velocity, elevation) can be included in the point spread function depending on the type of imaging sensor.
For illustration, \Cref{fig:sensor_measurement_trajectory} shows a sensor grid and measurement following this model.

% CUT-OFF PSF PARAMETER DESCRIPTION
%\text{C}(\state)$ denotes the cells of the measurement influenced by the presence of the object, and $\mathbbm{1}$ denotes the indicator function ($\mathbbm{1}_{\text{S}}(e) = 1 \text{ if } e \in \text{S}; 0 \text{ otherwise}$).

According to \eqref{eq:meas_model}, $\prob{\obss_{\tk}^{(j)}|\state_{\tk}}$ is Gaussian with mean $\widehat{\obss}^{(j)}$ and variance $\sigma_w^2$. Assuming that measurements among cells are conditionally independent, the complete likelihood is given by
\begin{equation}
\prob{\obs_\tk|\state_\tk} = \prod_{j=1}^\tJ \prob{\obss_\tk^{(j)}|\state_\tk}.
\label{eq:tbdlkl}
\end{equation}

\subsubsection{Riemann-Langevin proposal}

%\begin{equation}
%\log \prob{\obs_\tk|\state_\tk} = \sum_{j=1}^\tJ \log \prob{\obss_\tk^{(j)}|\state_\tk}.
%\end{equation}

The TBD likelihood in \eqref{eq:tbdlkl} consists of a product of $\tJ$ Gaussians. The log-gradient and information matrix of each of these Gaussians obey \eqref{eq:mvngradient} and \eqref{eq:mvnfisher}, respectively. Since the logarithm of the product becomes the sum of logarithms, we have that the TBD gradient and information matrix are
\begin{equation}
\nabla_{\state_\tk} \log \prob{\obs_\tk|\state_\tk} = \sum_{j=1}^\tJ \frac{1}{\sigma_w^2} \left(\frac{\partial \widehat{\obss}_\tk^{(j)}}{\partial \state_\tk}\right)\transpose (\obss_\tk^{(j)}-\widehat{\obss}_\tk^{(j)}),
\label{eq:tbdgradient}
\end{equation}
\begin{equation}
\Fisher{\state_{\tk}} = \sum_{j=1}^{\tJ} \frac{1}{\sigma_w^2} \left(\frac{\partial \widehat{\obss}_\tk^{(j)}}{\partial \state_\tk}\right)\transpose \left(\frac{\partial \widehat{\obss}_\tk^{(j)}}{\partial \state_\tk}\right).
\label{eq:tbdfisher}
\end{equation}
%
%In fact, due to the presence of $\mathbbm{1}_{\tC(\state_\tk)}(j)$ in $\widehat{\obss}_{\tk}^{(j)}$, $\widehat{\obss}_\tk^{(j)}$ is nonzero only when $j \in \tC(\state_\tk)$. Therefore,
%
%\begin{equation}
%\nabla_{\state_\tk} \log \prob{\obs_\tk|\state_\tk} = \sum_{j \in \tC(\state_\tk)} \frac{1}{\sigma_w^2} \left(\frac{\partial \widehat{\obss}_\tk^{(j)}}{\partial \state_\tk}\right)\transpose (\obss_\tk^{(j)}-\widehat{\obss}_\tk^{(j)}).
%\end{equation}

%For the purpose of illustration, let the state vector contain the two-dimensional position of an object,
%
%\begin{equation}
%  \state_\tk = \begin{bmatrix} x_\tk \\ y _\tk \end{bmatrix}
%\end{equation}
%
%then we have that
%
%\input{rlmcf_tbd_zhat_grad_eq}

\section{Riemann-Langevin Monte Carlo filter}
\label{sec:rlmcf}
Leveraging the results in the previous sections, including the adoption of a sequential MCMC approach and the Riemann-Langevin proposal, here we provide the full description of our \textit{Riemann-Langevin MC filter}. The details of all the steps required to implement our filter are given in Algorithm \ref{algo:rlmcf}. The algorithm has an outer-most loop in $\tk$ corresponding to the filtering time. The loop in $i$ represents the evolution of the Markov chain in the MCMC. Being a sequential MCMC, the MCMC contains two parts corresponding to the joint and refinement phases \cite{septier2016langevin}. Regarding the rest of the symbols introduced in Algorithm \ref{algo:rlmcf}: $\widehat{\text{p}}$ denotes the sample-based representation of a density; $\Nbi$ the burn-in length; and $\Np$ the number of particles.

\begin{algorithm}
\SetKwInOut{Input}{Input}\SetKwInOut{Output}{Output}
\Input{$\widehat{\text{p}}\left(\state_0\right)$, $\prob{\obs_{\tk}|\state_{\tk}}$, $\prob{\state_{\tk}|\state_{\tk-1}}, \obs_{1:\text{K}}$}
\Output{$\left\{\probhat{\state_{\tk}|\obs_{1:\tk}}\right\}_{\tk=1,\dots,\text{K}}$}
\BlankLine
\For{\upshape $\tk=1$ {\textbf{to} $\text{K}$}}{
  $\left(\state_{\tk},\state_{\tk-1}\right)^{0} \sim \prob{\state_{\tk}|\state_{\tk-1}}\widehat{\text{p}}\left(\state_{\tk-1}|\obs_{1:\tk-1}\right)$\; % Seed the Markov chain
  \For{$i=0$ {\upshape \textbf{to} $\Nbi+\Np-1$}}{
    \tcp{Joint draw.}
    $\left(\state_{\tk},\state_{\tk-1}\right)^{*} \sim \prob{\state_{\tk}|\state_{\tk-1}}\widehat{\text{p}}\left(\state_{\tk-1}|\obs_{1:\tk-1}\right)$\;
    $\AcceptanceRatioDefLkl$\;
    $u \sim \ContinuousUniform{0}{1}$\;
    \eIf{$u <$ {\upshape $\AcceptanceRatio$}}{
      $\left(\state_{\tk}, \state_{\tk-1}\right)^{i+1}=\left(\state_{\tk}, \state_{\tk-1}\right)^{*}$\;
    }{
      $\left(\state_{\tk}, \state_{\tk-1}\right)^{i+1}=\left(\state_{\tk}, \state_{\tk-1}\right)^{i}$\;
    }
    \tcp{Refinement.}
    $\state_{\tk}^{*} \sim \text{q}_{\text{RL}}\left({\state_{\tk}|\state_{\tk}^{i+1},\state_{\tk-1}^{i+1},\obs_{\tk}}\right)$\; % Sample proposal distribution
    $\AcceptanceRatioDefRL$\;
    $u \sim \ContinuousUniform{0}{1}$\;
    \If{$u <$ {\upshape $\AcceptanceRatioRef$}}{
      $\NewState = \ProposedState$\; % Acceptance, chain moves to proposed state 
    }
  }
  $\probhat{\state_{\tk}|\obs_{1:\tk}} = \Np^{-1} \sum_{i=\Nbi+1}^{\Nbi+\Np} \delta\left(\state_{\tk}-\state_{\tk}^{i}\right)$\; % Estimated posterior 
}
\caption{Riemann-Langevin MC filter.}
\label{algo:rlmcf}
\end{algorithm}

\section{Experiments}
\label{sec:experiment}

The performance of the Riemann-Langevin MC filter in \Cref{sec:rlmcf} leveraging second-order model knowledge is assessed in this section. In particular, we compare this filter to the traditional bootstrap PF and the standard sequential MCMC in a TBD application. We start by describing the tracking scenario (including the trajectory generation, motion and measurement models) and then the results are presented and analyzed.

The scenario is composed of a single object moving along a straight-line at a constant speed equal to 180 kilometers per hour. The object trajectory lasts 30 seconds and is shown in \Cref{fig:sensor_measurement_trajectory}. Range and bearing measurements are reported by an imaging sensor every second (i.e., sampling time $\Delta_t=1[s]$).

Unlike common high-dimensional experiments demonstrating the power of Hamiltonian/Langevin MC \cite{neal2011mcmc,girolami2011riemann,septier2016langevin}, our state space is only composed of four variables. Our motivation is to demonstrate that the Riemann-Langevin MC filter is not only useful in high-dimensions, but also in other circumstances challenging for PFs such as low-noise.

\subsection{Motion model}
\noindent
The state vector comprises two-dimensional (denoted by $x$ and $y$) position and velocity vectors:
\begin{equation}
\state_{\tk} = \begin{bmatrix} x_{\tk} & \dot{x}_{\tk} & y_{\tk} & \dot{y}_{\tk} \end{bmatrix}\transpose.
\end{equation}
The time evolution is given by a \textit{nearly constant velocity} model \cite{li2003survey} $\state_{\tk} = \text{A} \ \state_{\tk-1} + \mathbf{v}_{\tk-1}$, with transition matrix
\begin{equation}
\text{A} = \text{I}_{2} \otimes \begin{bmatrix} 1 & \Delta_t \\ 0 & 1 \end{bmatrix}
\end{equation}
where $\otimes$ denotes the Kronecker product. The noise  $\mathbf{v}_{\tk}$ is Gaussian with zero mean and covariance
\begin{equation}
\text{Q} = \begin{bmatrix} \sigma_{a_{x}}^2 & 0 \\ 0 & \sigma_{a_{y}}^2 \end{bmatrix} \otimes \begin{bmatrix} \Delta_t^3/3 & \Delta_t^2/2 \\ \Delta_t^2/2 & \Delta_t \end{bmatrix},
\end{equation}
where $\sigma_{a_{x}} = \sigma_{a_{y}} = 0.1 \ [\text{m} \ \text{s}^{-2}]$ denote scalar standard deviations of the acceleration along the $x$ and $y$ axes, respectively.

\subsection{Measurement model and sensor parameters}
\noindent
The TBD measurement model is described in \Cref{sec:tbd}. The specification of the parameters follows. The measurement noise is $\sigma_w = 10^{-4}$ and the $\text{SNR} = 80 \ [\text{dB}]$, simulating a very low-noise scenario especially challenging for PF. The sensor parameters are listed in \Cref{tab:sensor}. \Cref{fig:sensor_measurement_trajectory} depicts the sensor grid in a region of the field-of-view as well as a sample measurement.

\begin{table}
\centering
\caption{Sensor parameters}
\label{tab:sensor}
\begin{tabular}{c*{1}{|c}}
Range PSF constant (R) & $ 1.56 \cdot 10^6$ $[\text{m}^2]$ \\
Bearing PSF constant (B) & $1.88 \cdot 10^{-4}$ $[\text{rad}^2]$ \\
Range resolution & $500$ $[\text{m}]$ \\
Bearing resolution & $5 \cdot 10^{-3}$ $[\text{rad}]$ \\
Range lower bound & $22 \cdot 10^3$ $[\text{m}]$ \\
Range upper bound & $26 \cdot 10^3$ $[\text{m}]$ \\
Bearing lower bound & $-\pi/6$ $[\text{rad}]$ \\
Bearing upper bound & $\pi/6$ $[\text{rad}]$
\end{tabular}
\end{table}

\subsection{Tested methods and initialization}
We compare the following methods:
\begin{enumerate}
\item The Riemann-Langevin MC filter (\Cref{sec:rlmcf}).
\item The bootstrap PF based on IS and resampling \cite{gordon1993novel}.
\item Sequential MCMC with prior proposal \cite{septier2016langevin}.
\end{enumerate}

The Riemann-Langevin MC filter uses 400 particles, the bootstrap PF 5000, and the sequential MCMC with prior proposal 3000. For both MCMC methods the burn-in length is 100. Initial particles are drawn from two uniform distributions: for the position the area of the distribution is $1 \ [\text{km}^2]$; and for the velocity it is $100 \ [\text{m}^2 \ \text{s}^{-2}]$. Both distributions are centered around the ground truth.

\subsection{Numerical results}
\begin{figure}
\centering
\includegraphics[width=1.1\columnwidth]{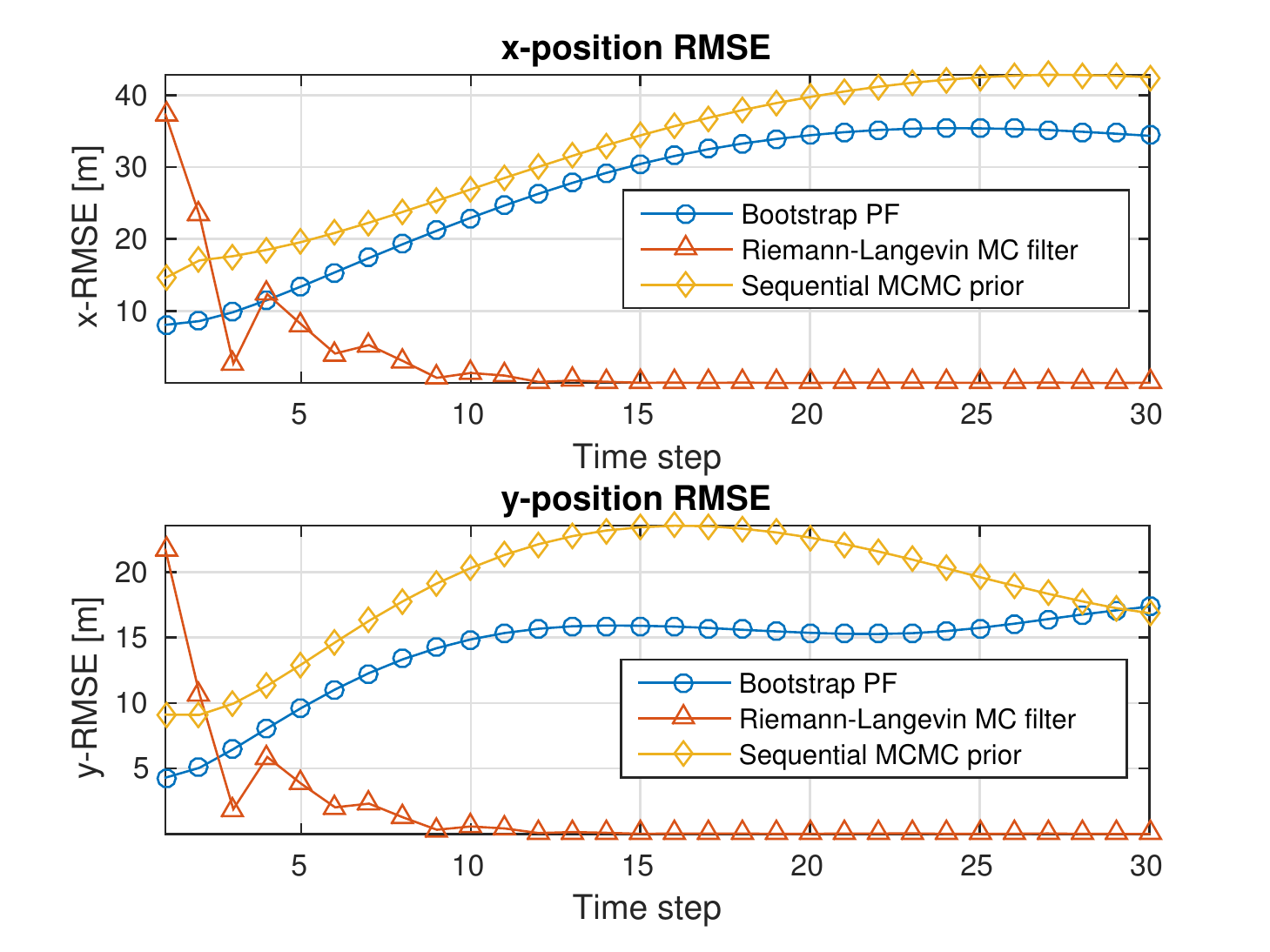}
\caption{Root mean squared error (RMSE) in the position state variables obtained with each of the methods in comparison.}
\label{fig:xyrmse}
\end{figure}

\noindent
The root mean squared error (RMSE) of the position estimate ($x$ and $y$ axes) is shown in \Cref{fig:xyrmse}. Results are obtained averaging $\nmcruns=50$ MC simulations. In all methods the estimated positions are given by the sample average over the population of particles (burn-in is discarded in the MCMC methods). The results show that the Riemann-Langevin MC filter outperforms the other methods despite its lower number of samples. Note that during the first few time steps the low performance of the Riemann-Langevin MC filter is due to its fewer number of particles. The superior overall performance stems from the Riemann-Langevin proposal, which leverages both prior and measurement information, whereas the proposals in the other methods only contain prior information.

In particular, the particle clouds in the Riemann-Langevin MC filter are diverse, whereas the particle clouds in the other methods are degenerated due to the low noise. The Riemann-Langevin proposal achieves diverse clouds via adaptation, focusing on the regions of the state space where most mass of the posterior density lies. A measure of the particles' degeneration (dispersion) with straightforward interpretation is the number of distinct particles. Across all the MC runs, the minimum and maximum number of distinct particles at the last time step are: 363 and 384 (of 400 total) in the Riemann-Langevin MC filter; 4 and 8 (of 5000) in the bootstrap PF; 4 and 11 (of 3000) in the sequential MCMC with prior proposal.

\section{Conclusion}
\label{sec:conclusion}
This letter presented the application of a  differential-geometric MCMC-based PF to the problem of TBD. The proposed Riemann-Langevin MC filter was, to the best of our knowledge, the first PF formulated for TBD that exploited measurement information in the proposal. The Riemann-Langevin proposal for TBD was presented along with the particular expressions for the TBD gradient and the Fisher information matrix. A low-dimensional experiment dealing with low-noise, a setup particularly challenging for PF, illustrated that the proposed filter outperformed the considered alternatives. In addition, the experiment demonstrated that the benefit of the Riemann-Langevin approach is not limited to high-dimensional filtering.

The application of the Riemann-Langevin MC filter is of course not limited to TBD. In this regard the Riemann-Langevin proposal and the associated algorithm  were formulated generically, enabling their application to a large class of filtering problems, with different transition and measurement models.

% or


% \appendices

% \section{}

%\section*{Acknowledgment}


\ifCLASSOPTIONcaptionsoff
  \newpage
\fi

\newpage
\IEEEtriggeratref{9}
\bibliography{../bibliography/bibliography}

\begin{thebibliography}{10}
\providecommand{\url}[1]{#1}
\csname url@samestyle\endcsname
\providecommand{\newblock}{\relax}
\providecommand{\bibinfo}[2]{#2}
\providecommand{\BIBentrySTDinterwordspacing}{\spaceskip=0pt\relax}
\providecommand{\BIBentryALTinterwordstretchfactor}{4}
\providecommand{\BIBentryALTinterwordspacing}{\spaceskip=\fontdimen2\font plus
\BIBentryALTinterwordstretchfactor\fontdimen3\font minus
  \fontdimen4\font\relax}
\providecommand{\BIBforeignlanguage}[2]{{%
\expandafter\ifx\csname l@#1\endcsname\relax
\typeout{** WARNING: IEEEtran.bst: No hyphenation pattern has been}%
\typeout{** loaded for the language `#1'. Using the pattern for}%
\typeout{** the default language instead.}%
\else
\language=\csname l@#1\endcsname
\fi
#2}}
\providecommand{\BIBdecl}{\relax}
\BIBdecl

\bibitem{gordon1993novel}
N.~J. Gordon, D.~J. Salmond, and A.~F. Smith, ``{Novel approach to
  nonlinear/non-Gaussian Bayesian state estimation},'' in \emph{IEE Proceedings
  on Radar and Signal Processing}, vol. 140, no.~2.\hskip 1em plus 0.5em minus
  0.4em\relax IET, 1993, pp. 107--113.

\bibitem{doucet2001sequential}
A.~Doucet, N.~de~Freitas, and N.~Gordon, \emph{{Sequential Monte Carlo methods
  in practice}}.\hskip 1em plus 0.5em minus 0.4em\relax Springer-Verlag, 2001.

\bibitem{arulampalam2002tutorial}
M.~S. Arulampalam, S.~Maskell, N.~Gordon, and T.~Clapp, ``{A tutorial on
  particle filters for online nonlinear/non-Gaussian Bayesian tracking},''
  \emph{IEEE Transactions on Signal Processing}, vol.~50, no.~2, pp. 174--188,
  2002.

\bibitem{khan2005mcmc}
Z.~Khan, T.~Balch, and F.~Dellaert, ``{MCMC-based particle filtering for
  tracking a variable number of interacting targets},'' \emph{IEEE Transactions
  on Pattern Analysis and Machine Intelligence}, vol.~27, no.~11, pp.
  1805--1819, 2005.

\bibitem{smith2008tracking}
K.~Smith, S.~O. Ba, J.-M. Odobez, and D.~Gatica-Perez, ``Tracking the visual
  focus of attention for a varying number of wandering people,'' \emph{IEEE
  Transactions on Pattern Analysis and Machine Intelligence}, vol.~30, no.~7,
  pp. 1212--1229, 2008.

\bibitem{pang2011detection}
S.~K. Pang, J.~Li, and S.~J. Godsill, ``Detection and tracking of coordinated
  groups,'' \emph{IEEE Transactions on Aerospace and Electronic Systems},
  vol.~47, no.~1, pp. 472--502, 2011.

\bibitem{bocquel2013multitarget}
M.~Bocquel, F.~Papi, M.~Podt, and H.~Driessen, ``{Multitarget tracking with
  multiscan knowledge exploitation using sequential MCMC sampling},''
  \emph{IEEE Journal of Selected Topics in Signal Processing}, vol.~7, no.~3,
  pp. 532--542, 2013.

\bibitem{salmond2001particle}
D.~Salmond and H.~Birch, ``A particle filter for track-before-detect,'' in
  \emph{Proceedings of the American Control Conference}, vol.~5.\hskip 1em plus
  0.5em minus 0.4em\relax IEEE, 2001, pp. 3755--3760.

\bibitem{boers2004multitarget}
Y.~Boers and J.~Driessen, ``Multitarget particle filter track before detect
  application,'' in \emph{IEE Proceedings Radar, Sonar and Navigation}, vol.
  151, no.~6.\hskip 1em plus 0.5em minus 0.4em\relax IET, 2004, pp. 351--357.

\bibitem{rutten2005recursive}
M.~G. Rutten, N.~J. Gordon, and S.~Maskell, ``Recursive track-before-detect
  with target amplitude fluctuations,'' in \emph{IEE Proceedings Radar, Sonar
  and Navigation}, vol. 152, no.~5.\hskip 1em plus 0.5em minus 0.4em\relax IET,
  2005, pp. 345--352.

\bibitem{duane1987}
S.~Duane, A.~D. Kennedy, B.~J. Pendleton, and D.~Roweth, ``{Hybrid Monte
  Carlo},'' \emph{Physics letters B}, 1987.

\bibitem{iglesias2015langevin}
F.~J. Iglesias~Garc{\'\i}a, M.~Bocquel, and H.~Driessen, ``{Langevin Monte
  Carlo filtering for target tracking},'' in \emph{18th International
  Conference on Information Fusion}.\hskip 1em plus 0.5em minus 0.4em\relax
  IEEE, 2015, pp. 82--89.

\bibitem{girolami2011riemann}
M.~Girolami and B.~Calderhead, ``{Riemann manifold Langevin and Hamiltonian
  Monte Carlo methods},'' \emph{Journal of the Royal Statistical Society:
  Series B (Statistical Methodology)}, vol.~73, no.~2, pp. 123--214, 2011.

\bibitem{iglesias2015acceptance}
F.~J. Iglesias~Garc\'{i}a, M.~Bocquel, P.~K. Mandal, and H.~Driessen,
  ``{Acceptance probability of IP-MCMC-PF: revisited},'' in \emph{10th Workshop
  on Sensor Data Fusion: Trends, Solutions, Applications}.\hskip 1em plus 0.5em
  minus 0.4em\relax IEEE, 2015.

\bibitem{septier2016langevin}
F.~Septier and G.~W. Peters, ``{Langevin and Hamiltonian based sequential MCMC
  for efficient Bayesian filtering in high-dimensional spaces},'' \emph{IEEE
  Journal of Selected Topics in Signal Processing}, vol.~10, no.~2, pp.
  312--327, 2016.

\bibitem{amari1985differential}
S.~Amari, \emph{Differential-Geometrical Methods in Statistics}.\hskip 1em plus
  0.5em minus 0.4em\relax Springer, 1985.

\bibitem{kay1993fundamentals}
S.~M. Kay, \emph{Fundamentals of Statistical Signal Processing, Volume I:
  Estimation Theory}.\hskip 1em plus 0.5em minus 0.4em\relax Prentice Hall,
  1993.

\bibitem{neal2011mcmc}
R.~M. Neal, ``{MCMC using Hamiltonian dynamics},'' in \emph{Handbook of Markov
  Chain Monte Carlo}.\hskip 1em plus 0.5em minus 0.4em\relax {Chapman \&
  Hall/CRC}, 2011.

\bibitem{li2003survey}
X.~R. Li and V.~P. Jilkov, ``{Survey of maneuvering target tracking. Part I.
  Dynamic models},'' \emph{IEEE Transactions on Aerospace and Electronic
  Systems}, vol.~39, no.~4, pp. 1333--1364, 2003.

\end{thebibliography}
\bibliographystyle{IEEEtran}

\end{document}